\documentclass[11pt,a4paper]{article}
\usepackage{jheppub}
\usepackage{graphicx}
\usepackage{slashed}
\usepackage{physics}
\usepackage{amsmath,amsbsy,amsfonts,amsmath,bbm,latexsym,amssymb,bm}
\usepackage{epsfig}
\usepackage{wasysym}
\usepackage{array}
\usepackage{mathrsfs}
\usepackage{dsfont}
\usepackage[super]{nth}
\newcolumntype{C}[1]{>{\centering\arraybackslash}m{#1}}

\newenvironment{Eqnarray}%
         {\arraycolsep 0.14em\begin{eqnarray}}{\end{eqnarray}}
\newcommand{\be}{\begin{equation}}
\newcommand{\ee}{\end{equation}}
\newcommand{\ba}{\begin{Eqnarray}}
\newcommand{\ea}{\end{Eqnarray}}
\newcommand{\bs}{\begin{subequations}}
\newcommand{\es}{\end{subequations}}

\newcommand{\grts}{\raise.3ex\hbox{$>$\kern-.75em\lower1ex\hbox{$\sim$}}}
\newcommand{\lets}{\raise.3ex\hbox{$<$\kern-.75em\lower1ex\hbox{$\sim$}}}

\def\eq#1{eq.~(\ref{#1})}
\def\eqst#1#2{eqs.~(\ref{#1})--(\ref{#2})}
\def\eqs#1#2{eqs.~(\ref{#1}) and (\ref{#2})}
\def\Eq#1{Eq.~(\ref{#1})}
\def\Eqs#1#2{Eqs.~(\ref{#1}) and (\ref{#2})}

\def\half{\tfrac12}

\def\ifmath#1{\relax\ifmmode #1\else $#1$\fi}
\def\lsub#1{\ifmath{_{\lower1.5pt\hbox{$\scriptstyle #1$}}}}
\def\lsup#1{^{\lower 6pt\hbox{$\scriptstyle#1$}}}
\def\phm{\phantom{-}}
\def\newcdot{\kern.06em{\cdot}\kern.06em}

\notoc

\allowdisplaybreaks

\begin{document}
\begin{flushright}
SCIPP-18/06
\end{flushright}

\title{Multi-Higgs doublet models: the Higgs-fermion couplings and their sum rules}

\author[a]{Miguel\ P.\ Bento}
\author[b]{Howard E.~Haber}
\author[a]{J.\ C.\ Rom\~{a}o}
\author[a]{Jo\~{a}o P.\ Silva}

\affiliation[a]{Centro de F\'{\i}sica Te\'{o}rica de Part\'{\i}culas (CFTP) and
    Departamento de F\'{\i}sica,
    Instituto Superior T\'{e}cnico, Technical University of Lisbon,
    1049-001 Lisboa, Portugal}

\affiliation[b]{Santa Cruz Institute for Particle Physics,
University of California,
Santa Cruz, California 95064, USA}
\emailAdd{miguelfilipebento@gmail.com}
\emailAdd{haber@scipp.ucsc.edu}
\emailAdd{jorge.romao@tecnico.ulisboa.pt}
\emailAdd{jpsilva@cftp.tecnico.ulisboa.pt}

\date{\today}

\abstract{This is the second of a series of papers that explores the physical parameterization, sum rules and unitarity bounds arising from a non-minimal scalar sector of the Standard Model (SM) that consists of $N$ Higgs doublets.  
In this paper, we focus on the structure and implication of the Yukawa interactions that couple the $N$ scalar doublets to the SM fermions.  We employ the charged Higgs basis, which is defined as the basis of scalar fields such that the neutral scalar field vacuum expectation value resides entirely in one of the $N$ scalar doublet fields, and the charged components of the remaining $N-1$ scalar doublet fields are the physical (mass-eigenstate) charged Higgs fields.  Based on the structure of the Yukawa Lagrangian of the model (and as a consequence of tree-level unitarity), one may deduce numerous sum rules, several of which have not appeared previously in the literature.  These sum rules 
can be used to uncover intimate relations between the structure of the Higgs-fermion couplings and the scalar/gauge couplings.  In particular, we show that the approximate alignment limit, in which the $W^+ W^-$ and $ZZ$ couplings to the observed Higgs boson are approximately SM-like, imposes significant constraints on the Higgs-fermion couplings.
}

\keywords{Higgs physics, Beyond Standard Model, Electroweak
interaction, CP violation, Discrete Symmetries}

\maketitle
\flushbottom

\section{Introduction}

The discovery of the Higgs boson at the Large Hadron Collider (LHC) at CERN
\cite{Aad:2012tfa, Chatrchyan:2012xdj, Khachatryan:2016vau}
leaves two crucial open questions.
First, how many elementary scalars there are in Nature?
Is there one single scalar as in the original proposal
for the Standard Model (SM),
or are there several scalar families, just like there are
several families of elementary fermions?
Second, are the
couplings of the observed 125 GeV scalar to gauge bosons and to fermions
consistent with the SM (and if yes, to what precision)?
Answering this second question will place important constraints
on models of new physics beyond the SM.
The two questions posed above are related.  For example,
a detailed study of the predictions of an $N$ Higgs doublet model
(NHDM) can guide experimental searches for new scalar phenomena.

Models with multiple scalar doublets are very rich. They
predict both neutral and charged scalars, whose mass basis do not,
in general, coincide with the interaction basis. As a result,
one has mixing among the neutral scalars and mixing among the
charged scalars.
In addition,
new sources of CP violation in the scalar sector are possible, and
the mechanism for CP violation may be spontaneous
\cite{Lee:1973iz} or explicit.
One might have new CP violation sources in the mixing of neutral CP-even and CP-odd scalars,
in the mixing of charged scalars,
and/or in the couplings of scalars with fermions.
In general, the NHDM also yields flavor changing
neutral scalar interactions, which are strongly constrained by experiment.
This difficulty is a challenge,
which one can address with extra symmetries,
such as the $\mathbb{Z}_2$ symmetry introduced in the two
Higgs doublet model (2HDM) following general theorems proposed by Glashow and Weinberg
\cite{Glashow:1976nt} and independently
by Paschos~\cite{Paschos:1976ay}.
However, the difficulty in avoiding flavor changing interactions mediated by neutral scalars is also an opportunity.
For example, the symmetries employed in suppressing flavor changing
neutral scalar interactions 
might also be related to the hierarchy of fermion
masses and mixing,
or even to the existence of Dark Matter.

In order to study models of extended Higgs sectors, one needs first to establish
a convenient notation and impose the relevant theoretical constraints,
such as those arising from unitarity bounds.
In a previous publication \cite{nosso},
we introduced a suitable notation in the pure scalar sector of the NHDM,
clearly related to the physical degrees of freedom, which
we identified as being those that appear in the mass basis
of the charged scalars,
and we studied the unitarity bounds arising from the scalar/gauge sectors.
Here, we will extend our parameterization into the fermion sector,
and we will study the unitarity bounds arising from interaction of
the fermions with the scalar/gauge sectors.

Section~\ref{sec:model} reviews and extends our physical parameterization and the
many relations among the parameters. These are used in
section~\ref{sec:sum} in order to derive new sum rules.
In section~\ref{sec:upper} we define a vector involving
the couplings in the gauge/scalar sector and
vectors involving the Yukawa couplings.
We show that the approximate Higgs alignment observed in the gauge/scalar sector
(where the properties of one neutral scalar is SM-like)
translates into an alignment between the gauge/scalar sector vector
and vectors in the scalar/fermion sector.
We present our conclusions in section~\ref{sec:conclusions}.
In appendix~\ref{app:sum},
we use the cancellation of bad high energy behavior
in $2 \rightarrow 2$ scattering amplitudes in order
to rederive the sum rules that have been obtained in
section~\ref{sec:sum} by looking directly at the Lagrangian.

\section{\label{sec:model} The $\bf\emph{N}$ Higgs doublet model with fermions}

In this section we discuss the
full Lagrangian of the most general
$N$ Higgs doublet model.
Our field content is the following:
(i) The usual SU(2)$_L \times$U(1)$_Y$ gauge bosons;
(ii) $N$ Higgs doublet fields, parameterized as:
\be
\Phi_k =
\left(
\begin{array}{c}
\varphi_k^+ \\
\tfrac{1}{\sqrt{2}} ( v_k + \varphi_k^0)
\end{array}
\right),
\label{Phi-1}
\ee
for $k=1,2,\ldots,N$; and
(iii) the quark doublets, $q_L=(p_L,n_L)$,
which is a vector in the $n_g$-dimensional family space
of left-handed doublets, and the quark singlets,
$p_R$ and $n_R$,
which are $n_g$-dimensional
vectors in the right-handed family spaces of charge
$2/3$ and $-1/3$ quarks, respectively.\footnote{Throughout
this paper we neglect leptons without loss of generality,
as our analysis of the quark sector is similar to
that of the lepton sector with Dirac neutrinos.}
The neutral scalar field vacuum expectation values are normalized such that
\be \label{vevnorm}
v^2\equiv v_1^2+v_2^2+\ldots v_N^2=(246~\rm{GeV})^2\,,
\ee
whose numerical value is fixed by the Fermi constant.

When expressed in terms of the physical gauge fields,
the covariant derivative may be written as
\be
i D_\mu
=
i \partial_\mu
- \frac{g}{2} (\tau_+ W^+_\mu + \tau_- W^-_\mu )
- e Q A_\mu
- \frac{g}{c_W}
\left( \frac{\tau_3}{2} - Q s_W^2 \right) Z_\mu,
\label{D_mu}
\ee
where $g$ is the SU(2) coupling constant,
$c_W = \cos{\theta_W}$,
$s_W = \sin{\theta_W}$,
$e$ is the positron charge,
$Q$ is the charge operator,
and the SU(2) generators, when acting on doublets,
\be
\tau_+ =
\left(
\begin{array}{cc}
0 &\quad \sqrt{2}\\
0 & \quad 0
\end{array}
\right),
\ \ \
\tau_- =
\left(
\begin{array}{cc}
0 & \quad 0 \\
\sqrt{2} & \quad  0
\end{array}
\right),
\ \ \
\tau_3 =
\left(
\begin{array}{cc}
1 & \,\,\,\phm 0 \\
0 & \,\,\, -1
\end{array}
\right).
\ee
The covariant derivative for the singlet right-handed quarks
only contains $i \partial_\mu$ and the terms proportional to
$Q$ in eq.~\eqref{D_mu}.
Our choice for the signs of the
coupling constants and of the gauge fields is that in
Ref.~\cite{Romao:2012pq} with all $\eta$ factors taken positive.
The kinetic terms are written as
\ba \label{kin_lag}
{\cal L}_{K \Phi}
&=&
(D_\mu \Phi_k)^\dagger (D_\mu \Phi_k),
\nonumber\\
{\cal L}_{K q}
&=&
\bar{q}_L (i \slashed{D}) q_L
+ \bar{p}_R (i \slashed{D}) p_R
+ \bar{n}_R (i \slashed{D}) n_R,
\ea
for the scalars and quarks, respectively.

For the scalar potential, we follow the notation
of \cite{BS,BLS}:
\be
V_H
=
\mu_{ij}  ( \Phi_i^\dagger \Phi_j )
+
\lambda_{ij,kl} ( \Phi_i^\dagger \Phi_j ) ( \Phi_k^\dagger \Phi_l )
= - {\cal L}_{\rm Higgs},
\label{VH}
\ee
where, by hermiticity,
\be
\mu_{ij} = \mu_{ji}^\ast,
\hspace{3ex}
\lambda_{ij,kl} \equiv \lambda_{kl,ij} = \lambda_{ji,lk}^\ast.
\label{hermiticity}
\ee

The Yukawa couplings are organized into
complex $n_g \times n_g$ matrices
$\Gamma_k$ and $\Delta_k$ (for $k=1,\ldots,N$),
as
\be
- {\cal L}_Y
= \bar{q}_L \Gamma_k \Phi_k n_R + \bar{q}_L \Delta_k \tilde{\Phi}_k p_R
+ \textrm{h.c.},
\ee
where $\tilde{\Phi}_k \equiv i \tau_2 \Phi_k^\ast$.

Under a weak basis transformation of the scalars\footnote{A weak basis
transformation is one which preserves the doublet structure.},
\be
\Phi_j = X_{j b} \Phi^\prime_b
\ee
the couplings with scalars transform into
\ba
\mu^\prime_{ab}
&=&
X_{i a}^\ast \mu_{i j} X_{j b},
\\
\lambda^\prime_{ab,cd}
&=&
X_{i a}^\ast X_{k c}^\ast \lambda_{i j, k l} X_{j b} X_{l d},
\\
\Gamma^\prime_b
&=&
\Gamma_j X_{jb},
\\
\Delta^\prime_b
&=&
\Delta_i X_{ib}^\ast.
\ea

In a previous publication \cite{nosso} we stressed the importance of the
charged Higgs basis, defined as the basis where the charged components
of all scalar doublets correspond to charged scalar mass eigenstates \cite{Nishi:2007nh}.
We may parameterize the fields in the charged Higgs basis as
\be \label{chhiggsbasis}
\Phi^C_1 =
\left(
\begin{array}{c}
G^+\\*[2mm]
\tfrac{1}{\sqrt{2}}
\left( v + H^0 + i G^0 \right)
\end{array}
\right),
\ \ \
\Phi^C_2 =
\left(
\begin{array}{c}
S^+_2\\*[2mm]
\tfrac{1}{\sqrt{2}} \varphi^{C0}_2
\end{array}
\right),
\ \ \
\ldots\,,
\ \ \
\Phi^C_N =
\left(
\begin{array}{c}
S^+_N\\*[2mm]
\tfrac{1}{\sqrt{2}} \varphi^{C0}_N
\end{array}
\right),
\ee
where  $S_1^+\equiv G^+$ is he charged massless would-be Goldstone boson and
$S_2^+,\ldots,S^+_N$ are the physical (mass-eigenstate)
charged Higgs fields, with corresponding masses $m^2_{\pm, a}$.
Notice that only the neutral component of the first doublet has a vacuum expectation value.
In Ref.~\cite{nosso}, it is shown that all scalar-scalar and scalar-gauge couplings
depend exclusively on a single $N \times 2N$ matrix $B$.
Its physical significance is the matrix that takes the neutral scalars
fields from the charged Higgs basis into their mass eigenstate basis.  Denoting $\varphi_1^{C0}\equiv H^0+iG^0$,
the neutral Higgs fields in the charged Higgs basis are given in terms of the neutral mass-eigenstate scalar fields,
\be
\varphi^{C0}_a
= \sum_{\beta=1}^{2N} B_{a \beta} S^0_\beta\, ,
\label{matrixB}
\ee
where $S_1^0\equiv G^0$ is the neutral massless would-be Goldstone boson and $S_2^0,\ldots,S^0_{2N}$ are the physical (mass-eigenstate)
neutral Higgs scalar fields, with corresponding masses $m^2_\beta$.
We may therefore introduce a basis transformation $X=U$, where $U$ is
a unitary matrix that diagonalizes the charged scalar squared-mass matrix such that
\begin{equation} \label{uka}
\varphi^+_k = \sum_{a=1}^{N} U_{k a} S^+_a \,,
\end{equation}
with the corresponding diagonal charged scalar squared-mass matrix denoted by
\be \label{Dpm2}
D_\pm^2 \equiv
\textrm{diag} \left( m_{\pm, 1}^2=0,  m_{\pm, 2}^2, \dots,  m_{\pm, N}^2 \right)\,,
\ee
and $U_{k1}=v_k/v$.

In the charged Higgs basis, the neutral fields $\varphi_a^{C0}$ are related to the neutral scalar fields of the original basis defined in \eq{Phi-1},
\be \label{UB}
\varphi_k^0=\sum_{a=1}^N U_{ka}\varphi_a^{C0}=\sum_{\beta=1}^{2N} \sum_{a=1}^N U_{ja}B_{a\beta}S^0_\beta\,.
\ee
Note that one can also diagonalize the neutral scalar squared-mass matrix starting from the original basis of scalar fields,
\be \label{Vdef}
\varphi^0_k= \sum_{\beta=1}^{2N} V_{k \beta} S^0_\beta \,,
\ee
with the corresponding diagonal neutral scalar squared-mass matrix denoted by
\be \label{D02}
D_0^2 \equiv
\textrm{diag} \left( m_1^2=0,  m_2^2, \dots,  m_{2N}^2 \right)\,,
\ee
and $V_{k1}=iv_k/v$.
\Eqs{UB}{Vdef} imply that,
\be \label{buv}
B=U^\dagger V\,.
\ee

It is straightforward to see that $U^\dagger$ is the
matrix that takes the scalar doublets from
the original basis to a charged Higgs basis.
Because the latter is defined up to a rephasing
of $N-1$ doublets \cite{nosso}, the transformation
into this basis is not unique. For example,
one could consider a matrix $U'=U K$ where~\!\footnote{Here, we shall assume that there are no mass degeneracies among the charged Higgs bosons.   If mass degeneracies exist, then the most general form for $K$ would be
a block diagonal form with an $n\times n$ unitary matrix replacing a diagonal matrix of phases within the $n$-dimensional mass-degenerate subspace.   For further details, see Ref.~\cite{Haber:2018iwr}.} 
\begin{equation}
K = \mathrm{diag}(1,e^{-i \chi_2},e^{-i \chi_3},\cdots,e^{-i \chi_N}) \, .
\end{equation}
Furthermore,
because $U$ is a basis transformation,
it is parameterized by $N^2$ non-physical
parameters. It is then easy to see that
the matrix
$B$ alone comprises all the relevant physical
parameters of the diagonalization of the charged
and neutral scalar fields.

The non-uniqueness of the charged Higgs basis implies that the matrix $U$ employed in \eq{uka} can be replaced
by $U^\prime =UK$.  That is
$U^\prime_{k1}=U_{k1}$ and $U^\prime_{ja}=e^{-i\chi_a}U_{ja}$ for $a=2,3,\ldots,N$.
\Eq{buv} then yields,
\be \label{BB}
B^\prime_{1\beta}=B_{1\beta}\,,\qquad\quad B^\prime_{a\beta}=e^{i\chi_a}B_{a\beta}\,,\quad \text{for $a=2,3,\ldots,N$}.
\ee
The unphysical phases $\chi_a$ can be absorbed into the definition of the charged Higgs basis scalar doublet 
fields, $\Phi_a^C$.  That is, 
\be \label{pac}
\Phi_a^C\to e^{i\chi_a}\Phi^C_a\,,\quad \text{for $a=2,3,\ldots,N$}.
\ee
In particular, note that under this rephasing, the charged Higgs fields rephase in the same way, i.e., $S_a^+\to e^{i\chi_a}S_a^+$ (for $a=2,3,\ldots,N$).  In contrast, the mass eigenstate neutral Higgs fields $S_\beta^0$ are invariant under this rephasing in light of \eq{UB} since the rephasing of $\phi_a^{C0}$ is consistent with the rephasing of $B_{a\beta}$. 

In defining the neutral Higgs mass eigenstate fields, $S_\beta^0$, one always has the option to redefine any of the neutral scalar fields via $S_\beta^0\to -S_\beta^0$ .   This means that the choice of the matrix elements of the diagonalization matrix in \eq{Vdef} is unique only up to a sign, $V_{k\beta}\to -V_{k\beta}$.  That is, one is free to change the overall sign of any column of $V$.   
For example, taking $h=S_2^0$  to be the 125 GeV neutral Higgs
field, the overall signs of the couplings,
$\textrm{sgn}[hVV]$ and $\textrm{sgn}[hbb]$,
do not have physical significance, although the relative sign of these two couplings is physical and can be measured.\footnote{This is the source of some
confusion in the literature, even in the 2HDM.}
It is common practice to employ a specific sign convention to uniquely fix the signs of the neutral Higgs mass eigenstate fields. In this convention, $V_{k1}=iv_k/v$ and the $V_{kj}$ (for $j=2,3,\ldots,2N$) are parameterized by $(N-1)(2N-1)$
real angles $\theta_{k\ell}$ (for $1\leq k<\ell\leq 2N-1$)~\cite{murnaghan}.   The ranges of the $\theta_{k\ell}$ can then be chosen to uniquely fix the signs of the columns of $V$~\cite{Haber:2006ue}.

In Ref.~\cite{nosso}, the following properties of the $N\times 2N$ matrix $V$ were obtained,
\ba 
\Re(V^\dagger V)&=&\mathds{1}_{2N\times 2N}\,,\label{vv1} \\
VV^\dagger &=& 2\newcdot  \mathds{1}_{N\times N}\,,\label{vv2} \\ 
VV^T&=&0\,.\label{vv3}
\ea
Several properties of $B$ have been
thoroughly studied in Ref.~\cite{nosso},
extending previous work in
refs.~\cite{Grimus:2002ux,Grimus:2007if,Grimus:2008nb,Grimus:1989pu}.
For example, in light of \eq{buv} and using the fact that $U$ is unitary, \eqst{vv1}{vv3} yield,
\ba
\tfrac12 B B^\dagger
&=& \mathds{1}_{N \times N}\,, \label{bbid1} \\
\Re (B^\dagger B)
&=& \mathds{1}_{2N \times 2N}\,,\label{bbid2} \\
BB^T &=& 0\,.\label{BBT}
\ea

We may also define a new orthogonal and antisymmetric matrix
\be \label{Adef}
A=\Im(B^\dagger B)\, ,
\ee
which appears in gauge boson couplings to two neutral scalars.  
Using \eq{bbid2}, one can write,
\be \label{BIA}
B^\dagger B=\mathds{1}_{2N \times 2N}+iA\,.
\ee
From \eq{Adef}, it immediately follows that, 
\be
A^T = - A\,,\qquad\quad A A^T = -A^2=\mathds{1}_{2N \times 2N}\,,
\ee
after employing \eqst{bbid1}{BBT}.
Furthermore, after multiplying \eq{BIA} on the left by $B$ and using \eq{bbid1}, we obtain
\be \label{eq:new_1}
BA=-iB\,,
\ee
\Eq{eq:new_1}, which is stated here explicitly for the first time, plays a significant role in the intermediate steps of the calculations employed later in this work.
Finally, we note the following explicit relations previously obtained in Ref.~\cite{nosso},
\ba
B_{a1} &=& i \delta_{a1},\label{properties_B1} \\
B_{1 \beta} &=& - A_{1 \beta} + i \delta_{1 \beta}\,.
\label{properties_B2}
\ea
\Eq{properties_B1} is just the statement that $S_1^0=G^0$ resides entirely in the imaginary part of $\varphi_1^{C0}$.
Note that \eq{properties_B2} can be obtained by using the relation,
$A_{1\beta}=-\Re B_{1\beta}$, which is a consequence of \eqs{Adef}{properties_B1}.

In the charged Higgs basis, the scalar potential takes the following form,
\be
\mathcal{V}_H
=
Y_{ab}  ( \Phi_a^{C \dagger} \Phi^C_b )
+
Z_{ab,cd} ( \Phi_a^{C \dagger} \Phi^C_b) ( \Phi_c^{C \dagger} \Phi^C_d).
\label{VH_chHbasis}
\ee
The minimization the scalar potential in the charged Higgs basis and the identification of the charged Higgs boson squared-masses are neatly summarized by the following equation obtained in Ref.~\cite{nosso},
\be \label{stationarity_CHB}
Y_{ab} + v^2 Z_{ab,11}
=(D_\pm^2)_{ab},
\ee
In addition,
there are a number of notable relations among the squared-masses of the neutral and charged scalars,
the physical mixing matrices ($A$ and $B$),
and the coefficients of the scalar potential in the charged Higgs basis ($Y$ and $Z$).  For example, performing the diagonalization of the neutral scalar squared-mass directly in the charged Higgs basis yields~\cite{nosso},
\ba
2 v^2 Z_{i1,1j}
&=&
-2 (D_\pm^2)_{ij} + \left( B D_0^2 B^\dagger\right)_{ij}\, ,
\label{eq:new_2}
\\
2 v^2 Z_{i1,j1}
&=&
\left( B D_0^2 B^T \right)_{ij}\,.
\label{eq:new_3}
\ea
Using these results along with \eqs{properties_B2}{stationarity_CHB}, one can easily derive,
\ba
Y_{1a}
&=&
\tfrac{1}{2} \left( A D_0^2 B^\dagger\right)_{1a}\, ,
\label{eq:new_4}
\\
Y_{a1}
&=&
- \tfrac{1}{2} \left( B D_0^2 A \right)_{a1}\, .
\label{eq:new_5}
\ea

In the charged Higgs basis, the Yukawa Lagrangian takes the following form,
\be
- {\cal L}_Y
= \bar{q}_L \Gamma_a^C \Phi_a^C n_R + \bar{q}_L \Delta_a^C \tilde{\Phi}^C_a p_R
+ \textrm{h.c.},
\ee
where
\be \label{phitilde}
\tilde{\Phi}_a^C
\equiv
\left(
\begin{array}{c}
\tfrac{1}{\sqrt{2}} (\varphi^{C0}_a)^\ast\\
- S_a^-
\end{array}
\right).
\ee
The quarks are brought into their mass basis by
unitary transformations
\ba
n_R
&=&
U_{dR}\, d_R,\qquad\quad
p_R=
U_{uR}\, u_R,
\\
n_L
&=&
U_{dL}\, d_L,
\qquad\quad
\,p_L
=
U_{uL}\, u_L.
\ea
Since only the neutral component of the first doublet $\Phi^C$ has a vacuum expectation,
these transformations are chosen such that
\ba
\frac{v}{\sqrt{2}}
U_{d L}^\dagger \Gamma_1^C U_{d R} = D_d
&=& \textrm{diag} (m_d, m_s, m_b, \dots), \label{DDdef}
\\
\frac{v}{\sqrt{2}}
U_{u L}^\dagger \Delta_1^C U_{u R} = D_u
&=& \textrm{diag} (m_u, m_c, m_t, \dots). \label{DUdef}
\ea
The Yukawa Lagrangian can then be rewritten as,
\ba
-\frac{v}{\sqrt{2}} {\cal L}_Y
&=&
\left(\bar{u}_L V, \bar{d}_L\right)
\left( D_d \Phi_1^C + N_d^{(2)} \Phi_2^C
    + \dots + N_d^{(N)} \Phi_N^C
\right) d_R
\nonumber\\
&&
+\,
\left(\bar{u}_L , \bar{d}_L V^\dagger\right)
\left( D_u \tilde{\Phi}_1^C + N_u^{(2)} \tilde{\Phi}_2^C
    + \dots + N_u^{(N)} \tilde{\Phi}_N^C
\right) u_R
+ \textrm{h.c.},
\label{Yuk_CHB}
\ea
where $V=U_{uL}^\dagger U_{dL}$ is the Cabibbo-Kobayashi-Maskawa
(CKM) matrix, and
\ba
N_d^{(a)}
&=&
\frac{v}{\sqrt{2}}
U_{d L}^\dagger \Gamma_a^C U_{d R} ,
\label{def_Nd}
\\
N_u^{(a)}
&=&
\frac{v}{\sqrt{2}}
U_{u L}^\dagger \Delta_a^C U_{u R},
\label{def_Nu}
\ea
with $a=2,3,\dots,N$.  In light of \eq{pac}, the matrices $N_d^{(a)}$ and $N_u^{(a)}$ rephase under the rephasing of the charged Higgs basis,
\be \label{NN}
N_d^{(a)}\to e^{-i\chi_a}N_d^{(a)}\,,\qquad\quad 
N_u^{(a)}\to e^{i\chi_a}N_u^{(a)}\,,\quad \text{for $a=2,3,\ldots,N$}.
\ee

In general,
the matrices $N_d^{(a)}$ and $N_u^{(a)}$ are not diagonal,
leading to flavor-changing neutral scalar interactions,
which are strongly constrained experimentally.
Notice that these matrices, multiplied by the appropriate
CKM matrix element, will also be responsible for the charged scalar interactions
with quarks.
Using \eqs{matrixB}{phitilde}, it follows that
\ba \label{yuk_lag}
-\frac{v}{\sqrt{2}} {\cal L}_Y
&=&
\bar{u}_L V \left( N_d^{(a)} S_a^+ \right) d_R
+\, \frac{v}{\sqrt{2}} \bar{d}_L D_d d_R
+ \frac{1}{\sqrt{2}} \bar{d}_L \left( N_d^{(a)} B_{a \beta} S_\beta^0\right) d_R
\label{yuks}
\\*[1mm]
&-&
 \bar{d}_L V^\dagger \left( N_u^{(a)} S_a^-\right) u_R
+\, \frac{v}{\sqrt{2}} \bar{u}_L D_u u_R
+ \frac{1}{\sqrt{2}} \bar{u}_L \left( N_u^{(a)} B_{a \beta}^\ast S_\beta^0\right) u_R
+ \textrm{h.c.}\,, \nonumber
\ea
where the sums over repeated indices run over all values of $a=1,\dots,N$
and $\beta=1,\dots,2N$, and  
\be \label{Ndone}
N_d^{(1)}=D_d\,,\qquad\quad N_u^{(1)}=D_u\,,
\ee
are identified as the diagonal down-type and up-type fermion mass matrices defined in \eqs{DDdef}{DUdef}.
Note that \eq{yuks} includes the fermion interactions with the charged and neutral Goldstone bosons, $S_1^+=G^+$ and $S_1^0=G^0$.  Moreover, \eq{yuks} is invariant under the rephasing of the charged Higgs basis in light of \eqs{BB}{NN}.


\begin{table}[tbp]
\centering
		\begin{tabular}{|c|c|c|c|}
		\hline
\multicolumn{2}{|c|}{Kinetic Lagrangian} &
\multicolumn{2}{c|}{Yukawa Lagrangian} \\
\hline
Coupling
& Feynman rule
& Coupling
&  Feynman rule  \\*[5pt]
\hline
\rule{0pt}{3ex}
$[\bar{u}_n d_m W^+_\mu]$
& $ - i \frac{g}{\sqrt{2}} \gamma_\mu P_L V_{n m} $
& $[ \bar{u}_n d_m G^+]$
& $ i \frac{g}{\sqrt{2}} \left( \frac{m^u_n}{M_W} P_L -
\frac{m^d_m}{M_W} P_R \right) V_{n m} $ \\*[5pt]
\hline
\rule{0pt}{3ex}
$[\bar{d}_n u_m W^-_\mu]$
& $ - i \frac{g}{\sqrt{2}} \gamma_\mu P_L V^*_{n m} $
& $[\bar{d}_n u_m G^-]$
&  $ i \frac{g}{\sqrt{2}} \left( \frac{m^u_n}{M_W} P_R -
\frac{m_m^d}{M_W} P_L \right) V^*_{n m} $ \\*[5pt]
\hline
\rule{0pt}{3ex}
$[\bar{u}_n u_m Z_\mu]$
& $ -i \frac{g}{c_w} \gamma_\mu \left( \frac{1}{2} P_L - \frac{2}{3} s_w^2 \right)
\delta_{n m} $
& $[\bar{u}_n u_m G^0]$
& $ -\frac{g}{2 M_W} m^u_n \delta_{n m} \gamma_5 $
\\*[5pt]
\hline
\rule{0pt}{3ex}
$[\bar{d}_n d_m Z_\mu]$
& $  i \frac{g}{c_w} \gamma_\mu \left(  \frac{1}{2} P_L - \frac{1}{3} s_w^2 \right)
\delta_{n m} $
& $ [\bar{d}_n d_m G^0] $
& $ \frac{g}{2 M_W} m^d_n \delta_{n m} \gamma_5 $
\\*[5pt]
\hline
		\end{tabular}
		\caption{\label{table:FD_cubic}
       The couplings of the up-type and down-type fermions to the massive gauge bosons and their Yukawa Lagrangian counterparts
        obtained by substituting the gauge bosons by the
        corresponding Goldstone bosons.}
\end{table}


The matrices $B$ and $N^{(a)}$
fully parameterize the Yukawa Lagrangian. Thus, we
may use the equivalence theorem~\cite{Veltman:1989ud} in order to relate
 some of the cubic couplings from the kinetic Lagrangian in
eq.~\eqref{kin_lag} to the couplings with Goldstone bosons
in eq.~\eqref{yuk_lag},
through the properties of $B$ in \eqs{properties_B1}{properties_B2}.
In contrast with our previous publication
\cite{nosso}, where both the scalar potential and the kinetic Lagrangian
Feynman rules were in general distinct from the SM, 
here we find that the fermion-gauge couplings of the NHDM
are identical to those of the SM. These results are
presented in table~\ref{table:FD_cubic},
where $m^u_n$ ($m^d_n$) is the $n$-th up-quark (down-quark) mass.
Thus, no new sum rules arise exclusively from the fermion-gauge couplings.

\section{\label{sec:sum}Sum rules}

A comprehensive study of sum rules for Higgs couplings in extended Higgs sectors (under the assumption of a CP-conservation) was first provided in Ref.~\cite{Gunion:1990kf}.   In Ref.~\cite{nosso}, we specialized to the NHDM (while relaxing the assumption of CP conservation in the scalar sector) and derived numerous sum rules involving the Higgs couplings in the scalar-gauge sector of the model  (see also Refs.~\cite{Grinstein:2013fia,Ginzburg:2015ata}).  In this section, we extend our study of the NHDM sum rules to include the Higgs couplings to fermions.   

We use the same notation of Ref.~\cite{nosso} in which
$\left[ X_a Y_b Z_c \right]$ is identified as the term in the Lagrangian
that depends explicitly on family type indices.
For example \cite{nosso},
from the Feynman rules
\ba
Z_\mu S_\beta^0 S_\gamma^0
&:&
\ 
\frac{g}{2 c_W}
(p_\beta^0 - p_\gamma^0)_\mu A_{\beta \gamma}\, ,
\nonumber\\*[1mm]
Z_\mu Z_\nu S_\beta^0
&:&
\ 
- \frac{i g M_Z}{c_W} A_{1 \beta}\, g_{\mu \nu}\, ,
\nonumber\\*[1mm]
W_\mu^+ W_\nu^- S_\beta^0
&:&
\ 
- i g M_W A_{1 \beta}\, g_{\mu \nu}\, ,
\nonumber\\*[1mm]
W^+ S^-_a S^0_\beta
&:&
\ 
\frac{i g}{2} (p_a^- - p_\beta^0)^\mu B_{a \beta}\, ,
\ea
we define
\ba
\left[ Z_\mu S_\beta^0 S_\gamma^0 \right]
&=&
A_{\beta \gamma}\, ,\qquad\quad
\left[Z_\mu Z_\nu S_\beta^0 \right]
=
A_{1 \beta}\, ,
\nonumber\\*[1mm]
\left[W_\mu^+ W_\nu^- S_\beta^0 \right]
&=&
A_{1 \beta}\, ,\qquad\quad
\left[W^+ S^-_a S^0_\beta\right]
=
B_{a \beta}\, .
\label{coups_VVS_VSS}
\ea
Since the matrix $A$ is antisymmetric, $A_{\beta \beta} = 0$
whenever the two indices coincide.
\clearpage

Analogously, we define
$\left[ X_a Y_b Z_c \right]_{R,L}$ as the term that depends on family type
indices that is proportional to the corresponding chiral projection operator $P_{R,L} \equiv \half(1 \pm \gamma_5)$.
For example,
in the Lagrangian term
\be
{\cal L} \supset C_1\, 
\left\{ f(a,b,c) P_L + g(a,b,c) P_R \right\} X_a Y_b Z_c,
\label{LandR_factors}
\ee
involving the fields $X_a$, $Y_b$, $Z_c$,
we identify
$[X_a Y_b Z_c]_{L}=f(a,b,c)$ and $[X_a Y_b Z_c]_{R}=g(a,b,c)$.
We employ indices $a$ and $\beta$ for scalars and
indices $m$, $n$, $p$ and $q$ as fermion family indices, and we follow closely the sign conventions of Ref.~\cite{Romao:2012pq}.
For convenience, we have extracted a normalization factor $C_1$, whose value depends on whether the scalar field
is electrically charged or neutral.
As an example, for the couplings of the charged scalars to fermion pairs,
it is convenient to define $C_1=\sqrt{2}/v=g /(\sqrt{2}\,m_W)$.  Then,\footnote{One can check that \eqst{udsL}{dusR} and \eqst{ddsL}{uusR} with the respective choices for $C_1$ are consistent by comparing these couplings for $\beta=1$ [cf.~\eq{properties_B1}] with the SM couplings of the charged and neutral Goldstone boson to corresponding quark-antiquark pairs~\cite{Gunion:1989we}. \label{fnv}} 
\ba
\left[ \bar{u}_n d_m S^+_a \right]_L &=& \left( N^{\dagger (a)}_u \right)_{n p} V_{p m} \, , \label{udsL}\\
\left[ \bar{u}_n d_m S^+_a \right]_R &=& -  V_{n p} \left(  N^{ (a) }_d \right)_{p m} \, , \\
\left[ \bar{d}_n u_m S^-_a \right]_L &=& - \left( N^{\dagger (a)}_d \right)_{n p}
(V^\dagger)_{p m} \, , \\
\left[ \bar{d}_n u_m S^-_a \right]_R &=&  (V^\dagger)_{n p}
\left(  N^{ (a) }_u \right)_{p m} \, , \label{dusR}
\ea
where repeated indices are summed.
For the couplings of the neutral scalars to fermion pairs,
it is convenient to define $C_1=-1/v=-g /(2\,m_W)$.  Then,\textsuperscript{\ref{fnv}}
\ba
\left[ \bar{d}_n d_m S^0_\beta \right]_L &=& \left( N^{\dagger (a)}_d \right)_{n m}
(B^\dagger)_{\beta a} \, ,\label{ddsL} \\
\left[ \bar{d}_n d_m S^0_\beta \right]_R &=& \left( N^{(a)}_d \right)_{n m}
B_{a \beta} \, , 
\label{ddS_R}
\\
\left[ \bar{u}_n u_m S^0_\beta \right]_L &=& \left( N^{\dagger (a)}_u \right)_{n m}
B_{a \beta} \, , \\
\left[ \bar{u}_n u_m S^0_\beta \right]_R &=& \left( N^{(a)}_u \right)_{n m}
(B^\dagger)_{\beta a} \,.\label{uusR}
\ea
In light of eqs.~\eqref{Yuk_CHB}-\eqref{def_Nu}, the matrices
$N^{(a)}_f$ and the couplings defined
here have dimensions of mass.  Once again, one can verify that all Yukawa interactions are independent of the rephasing of the charged Higgs basis (taking into account the corresponding rephasing of the charged Higgs fields, $S_a^\pm$).

Based on the structure of the Yukawa Lagrangian of the NHDM,
one may deduce several sum rules that have not appeared previously in the literature.
For example,
\ba
\sum_{\beta=1}^{2 N}
[\bar{f}_n f_m S^0_\beta]_L [\bar{f}_p f_q S^0_\beta]_L
&=&
0\, ,
\label{sum:0a}\\*[1mm]
\sum_{\beta=1}^{2 N}
[\bar{f}_n f_m S^0_\beta]_R [\bar{f}_p f_q S^0_\beta]_R
&=&
0\, ,
\label{sum:0b}
\ea
where $f= u$ (for up-quarks) or $f=d$ (for down-quarks).\footnote{The sum rules exhibited in \eqs{sum:0a}{sum:0b}, in the special case of $n=m=p=q$, have been obtained previously in Ref.~\cite{Ginzburg:2015ata}.}  

To derive the sum rules above, we provide details on one of the derivations.
\ba
\sum_{\beta=1}^{2N} 
[\bar{u}_n u_m S^0_\beta]_L [\bar{u}_p u_q S^0_\beta]_L
& & =
\sum_{\beta=1}^{2N}\, 
\sum_{a=1}^{N}\, 
\sum_{b=1}^{N}\, 
\left( N^{\dagger (a)}_u \right)_{n m}
B_{a \beta}\ 
\left( N^{\dagger (b)}_u \right)_{p q}
B_{b \beta}
\nonumber\\
& & =
\sum_{a=1}^{N}\, 
\sum_{b=1}^{N}\, 
\left( N^{\dagger (a)}_u \right)_{n m}
\left( N^{\dagger (b)}_u \right)_{p q}\ 
\sum_{\beta=1}^{2N}\, 
B_{a \beta}\ 
B_{b \beta}\nonumber\\
& & =
\sum_{a=1}^{N}\, 
\sum_{b=1}^{N}\, 
\left( N^{\dagger (a)}_u \right)_{n m}
\left( N^{\dagger (b)}_u \right)_{p q}\ 
(BB^T)_{ab}
= 0\, ,
\ea
where the last equality is a consequence of \eq{BBT}.
There are numerous other cases that yields a factor $B B^T$ as above and thus produce a similar sum rule.
For example, 
\ba
\sum_{\beta=1}^{2 N}
[\bar{u}_n u_m S^0_\beta]_L [\bar{d}_p d_q S^0_\beta]_R
&=&
0\, ,
\label{sum:1_11a}
\\*[1mm]
\sum_{\beta=1}^{2 N}
[\bar{u}_n u_m S^0_\beta]_R [\bar{d}_p d_q S^0_\beta]_L
&=&
0\, .
\label{sum:1_11b}
\ea
Furthermore,
\ba
\sum_{q=1}^3
\left[ \bar{d}_n u_q S^-_b \right]_L
\left[ \bar{u}_q d_m S^+_a \right]_R
&=&
\left( 
 N^{\dagger (b)}_d\,  N^{(a)}_d
\right)_{n m}\, ,
\label{sum:2}
\\
\sum_{a=1}^N\, \sum_{q=1}^3
\left[ \bar{d}_n u_q S^-_a \right]_L
\left[ \bar{u}_q d_m S^+_a \right]_R
&=&
\frac{1}{2}
\sum_{\beta=1}^{2N}\, \sum_{q=1}^3
\left[ \bar{d}_n d_q S^0_\beta \right]_L
\left[ \bar{d}_q d_m S^0_\beta \right]_R
=
\sum_{a=1}^N
\left( 
 N^{\dagger (a)}_d\,  N^{(a)}_d
\right)_{n m}
\nonumber
\\
\phantom{line}
\label{sum:3}
\\
\sum_{a=1}^N\, \sum_{q=1}^3
\left[ \bar{u}_n d_q S^+_a \right]_L
\left[ \bar{d}_q u_m S^-_a \right]_R
&=&
\frac{1}{2}
\sum_{\beta=1}^{2N}\, \sum_{q=1}^3
\left[ \bar{u}_n u_q S^0_\beta \right]_L
\left[ \bar{u}_q u_m S^0_\beta \right]_R=
\sum_{a=1}^N
\left( 
 N^{\dagger (a)}_u\,  N^{(a)}_u
\right)_{n m}
\nonumber\\
\phantom{line}
\label{sum:4}
\ea

Combining the parameterization of the Yukawa Lagrangian presented here
with the parameterization of the scalar sector in Ref.~\cite{nosso},
we find
\begin{eqnarray}
\label{sum:1}
\sum_{\beta=1}^{2N} [\bar{f}_n f_m S^0_\beta]_{L} [Z_\mu S^0_\beta S^0_\alpha] &=& -i
[\bar{f}_n f_m S^0_\alpha]_{L} \, , \\
\sum_{\beta=1}^{2N} [Z_\mu S^0_\alpha S^0_\beta] [\bar{f}_n f_m S^0_\beta]_{R}  &=& -i
[\bar{f}_n f_m S^0_\alpha]_{R} \, .
\end{eqnarray}
We also observe that
\begin{eqnarray}
\sum_{\beta=1}^{2N} [\bar{f}_n f_m S^0_\beta]_{L} [V_\mu V_\nu S^0_\beta] &=& -
\left( D_f \right)_{n m} \, , \\
\sum_{\beta=1}^{2N} [V_\mu V_\nu S^0_\beta] [\bar{f}_n f_m S^0_\beta]_{R}  &=& -
\left( D_f \right)_{n m} \, ,
\end{eqnarray}
where,
as before,
$V_\mu V_\nu = Z_\mu Z_\nu , \, W_\mu^+ W_\nu^-$ and $f=u, d$.
We find it useful to write certain sum rules that arise from the fact
that the CKM matrix is unitary.
For example,
\begin{eqnarray}
\sum_{p,a} [\bar{u}_n d_p S^+_a]_L [W^+_\mu S^0_\beta S^-_a]
( V^\dagger )_{p m} &=& [\bar{u}_n u_m S^0_\beta]_L \, , \\*[1mm]
- \sum_{p,a} ( V^\dagger )_{n p} [\bar{u}_p d_m S^+_a]_R [W^+_\mu S^0_\beta S^-_a]
 &=& [\bar{d}_n d_m S^0_\beta]_R\, ,\\
\sum_{p,a} V_{n p} [\bar{d}_p u_m S^-_a]_R [W^-_\mu S^0_\beta S^+_a]
&=& [\bar{u}_n u_m S^0_\beta]_R \, , \\*[1mm]
- \sum_{p,a}  [\bar{d}_n u_p S^-_a]_L [W^-_\mu S^0_\beta S^+_a] V_{p m}
&=& [\bar{d}_n d_m S^0_\beta]_L \, .
\label{sum:last}
\end{eqnarray}

We have derived the sum rules above directly from the Lagrangian.
One can also obtain these sum rules from unitarity bounds.
Some sum rules were written for a general model in
eq.~(3.4) and eq.~(3.7) of Ref.~\cite{Gunion:1990kf}.
In appendix~\ref{app:sum} we show explicitly the derivation
of those sum rules based on the cancellation of bad high energy behavior in $2\to 2$ scattering processes,
in the case of the most general NHDM with fermions.
Note that
in contrast to the results of section V of Ref.~\cite{Gunion:1990kf},
the sum rules exhibited in table~\ref{table:FD_cubic} and in eqs.~\eqref{sum:1}--\eqref{sum:last}
have been derived under the assumption of 
multiple quark family generations.

\section{\label{sec:upper}A critical constraint from perturbativity}

The sum rules obtained in Section~\ref{sec:sum} can be used to uncover
intimate relations between the structure of Yukawa couplings
and the scalar/gauge couplings.
As an illustration, we start by observing that eq.~\eqref{BIA}
can be rewritten as,
\begin{equation}
\delta_{\beta \gamma} + i\, A_{\beta \gamma}
=
\sum_{a=1}^N B_{a\beta}^\ast B_{a\gamma}\, .
\end{equation}
Setting $\beta=\gamma = 2$ and noting that the matrix $A$ is antisymmetric,
we get
\begin{equation}
1 = \sum_{a=1}^N |B_{a 2}|^2\, .
\end{equation}
Thus,
$|B_{12}|$ must be smaller than one.  Moreover,
we know from eqs.~(\ref{properties_B1}), (\ref{properties_B2}) and (\ref{coups_VVS_VSS})
that $B_{12}=-[VV h_{125}]$,
where we have assumed that the lowest lying neutral scalar coincides
with the one found with 125 GeV at LHC
(recall that $a=1$ refers to the neutral would-be Goldstone
boson, while $a=2$ refers to the lowest lying massive neutral
mass eigenstate).
Therefore, one may parameterize
\begin{equation}
\left|[VV h_{125}]\right|^2=|B_{12}|^2 = s^2_{\bar\beta-\bar\alpha},
\label{def_b-a}
\end{equation}
where, henceforth $s_\theta$, $c_\theta$, and $t_\theta$
represent the sine, cosine, and tangent of any angle $\theta$
that appears in the subscript.\footnote{We recall in passing that this discussion implies
that the coupling $[VV h_{125}]$ measured at LHC is smaller
than unity in any multi-Higgs doublet model.
Had $[VV h_{125}]$ been found experimentally to be larger than one, then
not only the SM but any NHDM would have been excluded.}
Although reminiscent of the notation in the real 2HDM,
the definition of $s^2_{\bar\beta-\bar\alpha}$ in eq.~\eqref{def_b-a}
is completely general.
Since the value of $[VV h_{125}]$ deduced from the LHC Higgs data is very close to one,
we conclude that $c^2_{\bar\beta-\bar\alpha}$ is close to zero.

Let us now define
\begin{equation}
\vec{b} = \left[ B_{22}, B_{32}, \dots, B_{N2}\right]\, .
\end{equation}
Clearly, the squared-magnitude of the this vector, $|\vec{b}|^2 = c^2_{\bar\beta-\bar\alpha}$,
%
%
must be very close to zero.
Next, we recall from eq.~\eqref{matrixB} 
that the matrix $B$ takes the neutral scalars
fields from the charged Higgs basis into their mass basis.
Thus,
$s^2_{\bar\beta-\bar\alpha} \sim 1$ means that the massive neutral scalar
in the first doublet of the charged Higgs basis approximately coincides
with the lightest neutral scalar mass eigenstates, which is identified with the observed
Higgs boson of mass 125 GeV.
This is known as the alignment limit~\cite{Gunion:2002zf,Craig:2013hca,Haber:2013mia,Asner:2013psa,Carena:2013ooa,Carena:2014nza,Dev:2014yca,Bernon:2015qea,Bernon:2015wef}.
It occurs naturally in the decoupling limit \cite{Gunion:2002zf},
but can also arise in a parameter regime 
without decoupling.

We will now show that,
as a consequence of approximate alignment as suggested by the precision Higgs data,
the vector $\vec{b}$ -- which depends exclusively on properties of the neutral scalars --
must be almost orthogonal to the vectors
\begin{equation} \label{anm}
\vec{a}^{(nm)}
=
\left[
\left( N_d^{(2)} \right)_{nm},
\left( N_d^{(3)} \right)_{nm},
\dots,
\left( N_d^{(N)} \right)_{nm}
\right]\, ,
\end{equation}
for any choice of $m$ and $n$ (explicit reference to $n$ and $m$ will henceforth be suppressed).
Indeed, in light of the Cauchy-Schwarz inequality,
\begin{equation}
|\vec{a}\newcdot\vec{b}\,|^2
\leq
|\vec{a}|^2\, |\vec{b}|^2
= |\vec{a}|^2\, c^2_{\bar\beta-\bar\alpha}
\end{equation}
is suppressed by $c^2_{\bar\beta-\bar\alpha}$.
Eq.~\eqref{yuk_lag} shows that entries in the $\sqrt{2} N_f^{(a)}/v$
matrix are physical [up to an overall rephasing as shown in \eq{NN}], for they appear in the Yukawa Lagrangian expressed in terms of the
scalars fields in their mass basis.
Moreover,
for the theory to remain perturbative, such couplings cannot
exceed some reference value, which we take to be $4 \pi$.
As a result
\begin{equation}
|\vec{a}|^2 \leq \sum_{a \geq 2}
\left| \left( N_d^{(a)} \right)_{nm} \right|^2
\leq 8\pi^2 v^2 (N-1) \, ,
\end{equation}
and
\begin{equation}
|\vec{a}\newcdot \vec{b}\,|^2
\leq
8\pi^2 v^2 (N-1) \, c^2_{\bar\beta-\bar\alpha}.
\label{crucial_1}
\end{equation}
This shows that the alignment limit,
which is initially defined based on
the observed $VV h_{125}$ coupling,
has a dramatic impact on the Higgs-fermion Yukawa couplings.
This is one of our major results.
It can be written in a more interesting fashion by
setting $\beta=2$ in eq.~\eqref{ddS_R},
\begin{equation}
\left[ \bar{d}_n d_m S^0_2 \right]_R =
\left( N_d^{1}\right)_{nm} B_{12}
+
\sum_{a \geq 2}
\left( N^{(a)}_d \right)_{nm}
B_{a 2} \, .
\label{ddS2_eq1}
\end{equation}
In light of \eq{Ndone}, it follows that,
\begin{equation}
\left[ \bar{d}_n d_m h_{125} \right]_R
- m^d_n\, \delta_{nm}\, B_{12} = \vec{a}\newcdot\vec{b}
\label{ddS2_eq2}
\end{equation}
is also bounded by eq.~\eqref{crucial_1}.  Likewise, \eq{ddsL} yields
\begin{equation}
\left[ \bar{d}_n d_m h_{125} \right]_L
- m^d_n\, \delta_{nm}\, B^*_{12} = (\vec{a}^T\newcdot\vec{b})^*\,,
\label{ddS2_eq3}
\end{equation}
where $\vec{a}^{T(nm)}=\vec{a}^{(mn)}$ [cf.~\eq{anm}].
Similar equations hold for the up type fermions.
We conclude that the couplings of the observed 125 GeV scalar to quark pairs are approximately diagonal
in the alignment limit with values that approximate the corresponding coupling of the SM Higgs boson.
Moreover, the magnitude of the off-diagonal couplings of the 125 GeV scalar are bounded according to \eq{crucial_1}. Of course, this behavior is expected since the tree-level properties of $H^0\equiv\sqrt{2}\Re\Phi_1^{C}-v$ are precisely those of the SM Higgs boson.  In the alignment limit, $H^0$ is an approximate mass eigenstate that is identified as the observed 125 GeV scalar.

It is instructive to apply eq.~\eqref{ddS2_eq2} in the case of 
the so-called complex two Higgs doublet model (C2HDM) 
(see, e.g., refs.~\cite{Ginzburg:2002wt, ElKaffas:2007rq, Arhrib:2010ju,
Barroso:2012wz, Fontes:2014xva, Fontes:2015mea, Fontes:2015xva}).
A recent analysis was performed in \cite{Fontes:2017zfn},
introducing the public C2HDM\_HDECAY code for the
HDECAY program \cite{Djouadi:1997yw},
as well as all the corresponding Feynman rules
\cite{WebPageC2HDM}.
Using eq.~(B.12) of Ref.~\cite{nosso},
we find
\begin{equation}
\left[ \bar{d}_n d_m h_{125} \right]_R
+ m^d_n\, \delta_{nm}\, s_{\beta - \alpha}\, c_2 =
\left( N_d^{(2)}\right)_{nm} (- c_{\beta - \alpha} c_2 + i s_2).
\label{ddS2_C2HDM}
\end{equation}
In the C2HDM there are three mixing angles
($\alpha_1$, $\alpha_2$, and $\alpha_3$);
$c_2 \equiv \cos{\alpha_2}$ and similarly for others;
while we define $\alpha_1 = \alpha + \pi/2$,
in order to make contact between $\alpha_1$ as employed in the C2HDM
and the angle $\alpha$ used in its real 2HDM limit.
In the notation used here, the $h_{125}VV$ coupling
is given by $c_2 \cos{(\alpha_1 - \beta)} = - c_2\, s_{\beta - \alpha}$,
which corresponds to $s_{\bar\beta - \bar\alpha}$ used in eq.~\eqref{def_b-a}.
For the $h_{125}VV$ coupling to be close to unity,
$s_2$ must be close to zero (\textit{i.e.}, a small CP-violating angle),
and $c_{\beta - \alpha}$ must also be close to zero,
making both terms on the right-hand-side of eq.~\eqref{ddS2_C2HDM}
close to zero.  Consequently, the real part of
$\left[ \bar{d}_n d_m h_{125} \right]_R$ must lie close to its SM value
and its imaginary part must be close to zero.\footnote{Despite the bounds on $s_2$ and contrary to popular belief,
one can still have dominant CP-violating couplings
to some fermions,
even when the bounds from electric dipole moments are taken into account ~\cite{Fontes:2015mea, Fontes:2015xva, Fontes:2017zfn}.}
\Eq{ddS2_C2HDM} for the C2HDM,
and more generally eq.~\eqref{ddS2_eq2} in the case of the NHDM
can also be used to generalize the results presented recently
in Ref.~\cite{Dery:2017axi}.

\section{\label{sec:conclusions}Conclusions}

Although the Standard Model employs a Higgs sector consisting of a hypercharge-one, doublet of scalar fields, the generational structure of the fermionic sector invites us to consider the possibility that the Higgs sector of the Standard Model is also non-minimal, consisting of $N$ Higgs doublets.  Without prior knowledge of $N$, it is useful to analyze the NHDM in the case of general $N$.  In a previous paper~\cite{nosso}, we examined the physical parameterization, sum rules and unitarity bounds of the bosonic sector of the NHDM.  We were able to provide an elegant formulation of the NHDM by exploiting the charged Higgs basis, where the neutral scalar field vacuum expectation value resides entirely in one of the $N$ scalar doublet fields, and each of the remaining $N -1$ scalar doublet fields contains a physical (mass-eigenstate) charged Higgs field.  In this formulation, many of the purely bosonic couplings of the model can be expressed entirely in terms of an $N \times 2N$ matrix $B$.

This paper extends the results of Ref.~\cite{nosso} to include the most general Higgs-fermion Yukawa couplings.  We have shown that in addition to $B$, one must introduce a pair of $N-1$ complex $3\times 3$ matrices (one for up-type and one for down-type), along with the diagonal up and down-type quark mass matrices in order to fully parameterize the Higgs-quark Yukawa interactions.  Using these parameters, we have derived a set of sum rules that involve the Higgs-fermion interactions.  Some of these sum rules exclusively involve the Yukawa couplings, whereas others involve products of Yukawa couplings and gauge/Higgs couplings.  Several of these sum rules have not appeared previously in the literature.

In the charged Higgs basis, the tree-level couplings of the neutral CP-even component of the scalar doublet [denoted by $H^0$ in \eq{chhiggsbasis}] that contains the entire neutral scalar field vacuum expectation value correspond precisely to those of the SM Higgs boson.
In general, $H^0$ is not a mass-eigenstate due to the mixing of this field with the other neutral scalar fields of the NHDM.   However, if $H^0$ is an approximate mass eigenstate, then the Higgs sector is said to exhibit approximate alignment, since the corresponding mass eigenstate is approximately aligned in field space with the neutral Higgs vacuum expectation value.  The alignment limit can be conveniently defined by exploiting the sum rule satisfied by the $VV$ couplings to the neutral scalars (where $VV=W^+ W^-$ or $ZZ$).  We are then able to show the corresponding impact of the alignment limit on the Higgs-fermion couplings.

Of course, the sum rules governing the Higgs-fermion couplings of the NHDM, while constraining the model in interesting ways, do not address the phenomenological challenge presented by the near absence of flavor-changing neutral currents in the experimental data.   Without further model constraints, either via fine-tuning of couplings or by the imposition of additional symmetries, the generic NHDM will exhibit significant tree-level flavor changing neutral currents mediated by neutral Higgs exchange, in conflict with experimental observations.   Addressing this challenge will be the subject of a future publication.

\vspace{2ex}

\acknowledgments
The work of M.P.B, J.C.R. and J.P.S. is supported in part
by the Portuguese \textit{Funda\c{c}\~{a}o para a Ci\^{e}ncia e Tecnologia}
(FCT) under contracts CERN/FIS-NUC/0010/2015, PTDC/FIS-PAR/29436/2017 and UID/FIS/00777/2013.
H.E.H. is supported in part by the U.S. Department of Energy grant
number DE-SC0010107, and in part by the grant H2020-MSCA-RISE-2014
No. 645722 (NonMinimalHiggs).

\bigskip\bigskip

\centerline{\bf\Large Appendices}

\appendix

\section{ \label{app:sum} Generalized sum rules}

\subsection{Notation and Conventions}

In order to obtain the sum
rules of section III of Ref.~\cite{Gunion:1990kf},
in particular their equations (3.3),
(3.4) and (3.7),  it is convenient to adopt their conventions for the
Feynman rules, 
\begin{align}
  \label{eq:1}
&V^\alpha_a V^\beta_b V^\gamma_c :\ i\, g_{abc}\left[ (p_a -p_b)^\gamma
  + (p_b -p_c)^\alpha + (p_c -p_a)^\beta\right]
\equiv i\, g_{abc}\ \Gamma^{\alpha\beta\gamma}(p_a,p_b,p_c)
\\[+2mm]
&V^\alpha_a V^\beta_c \phi_i:\ i\, g_{abi}\ g^{\alpha\beta} \\[+2mm]
&V_a^\alpha \phi_i \phi_j: i\, g_{aij}\ (p_i-p_j)^\alpha \\[+2mm]
& V_a^\alpha \overline{f}_m f_n:\ i\, \gamma^\alpha \left(g^L_{amn} P_L +
  g^R_{amn} P_R\right)  \\[+2mm]
& \phi_i  \overline{f}_m f_n:\ i\, \left(g^L_{imn} P_L +
  g^R_{imn} P_R\right)
\end{align}
with all momenta incoming.
Here $f$, $V$, and $\phi$ stand for fermions, gauge bosons, and
scalars, respectively, and $P_{R,L}\equiv \half (1\pm\gamma_5)$ are the usual chiral projection operators.
We will use lowercase $m_n$ for the mass of the fermion $f_n$,
and uppercase $M_a$ for the mass of the gauge boson $V_a$.

\subsection{$FFVV$ Sum Rules}

\subsubsection{The amplitudes}

\begin{figure}[t!]
\center
\includegraphics[scale=1.1]{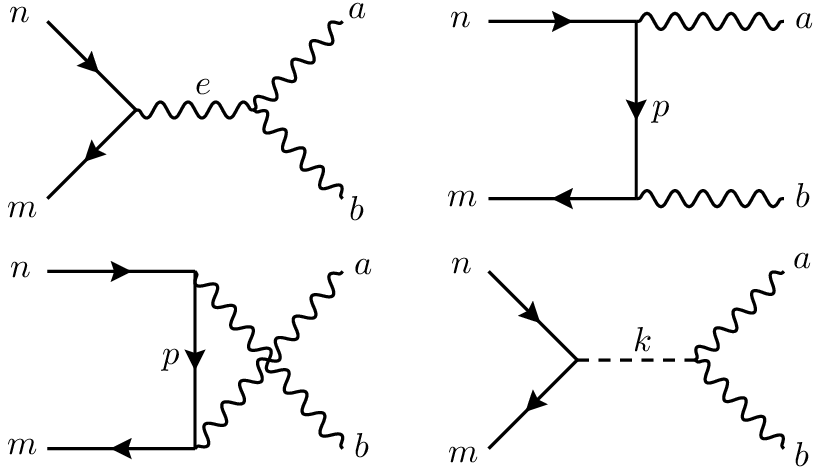}
  \caption{Diagrams for the scattering $f_n + \overline{f}_m \to V_a V_b$.}
  \label{fig:FFAA}
\end{figure}

The diagrams contributing to the scattering $f_n(p_1) + \overline{f}_m(p_2) \to
V_a(p_3) + V_b(p_4)$
are exhibited in fig.~\ref{fig:FFAA}.
In an obvious notation we will name the amplitudes according to
Mandelstam variables channel ($s,t$ or $u$) and by the particle being
exchanged. We then obtain,
\begin{align}
  \mathcal{M}_s^{A} =&(-i) (i\, g_{abe})
 \Gamma_{\alpha\beta\nu}(-p_4,-p_3,p_1+p_2)\ i\,
 \overline{f}_m(p_2)\gamma_\mu \left( g^L_{emn} P_L + g^R_{emn}
  P_R\right)\, f_n(p_1)
\nonumber\\[6pt]
&\hskip 5mm
\times (-i) \frac{\left[g^{\mu\nu} -
(p_1+p_2)^\mu (p_1+p_2)^\nu/M_e^2 \right]}
{s -M_e^2}
\epsilon^\alpha(p_3) \epsilon^\beta(p_4)\, ,\nonumber \\
 \mathcal{M}_t^{f} =&(-i) (i)^3\,
\overline{f}_m(p_2)\gamma_\beta \left( g^L_{bmp} P_L + g^R_{bmp}
P_R\right) (\slashed{p}_1-\slashed{p}_3 + m_p) \gamma_\alpha \left(
g^L_{apn} P_L + g^R_{apn} P_R\right)  f_n(p_1)
\nonumber\\
&\hskip 5mm
\times \frac{1}{t -m_p^2}
\epsilon^\alpha(p_3) \epsilon^\beta(p_4)\, ,
\nonumber\\
  \mathcal{M}_u^{f} =&(-i) (i)^3\,
\overline{f}_m(p_2)\gamma_\alpha \left( g^L_{amp} P_L + g^R_{amp}
P_R\right) (\slashed{p}_1-\slashed{p}_3 + m_p) \gamma_\beta \left(
g^L_{bpn} P_L + g^R_{bpn} P_R\right)  f_n(p_1) \nonumber\\
&\hskip 5mm
\times\frac{1}{u -m_p^2}
\epsilon^\alpha(p_3) \epsilon^\beta(p_4)\, ,
\nonumber\\
  \mathcal{M}_s^{\phi} =& (-i) (i)^3 g_{abk}\,
\overline{f}_m(p_2)\left( g^L_{kmn} P_L + g^R_{kmn}
  P_R\right)\, f_n(p_1)
     \frac{g_{\alpha\beta}}{s-m_k^2}
\epsilon^\alpha(p_3) \epsilon^\beta(p_4)\, .
\end{align}

\subsubsection{The high energy limit}

When the gauge bosons are longitudinally polarized the diagrams of
fig.~\ref{fig:FFAA} grow with energy for large center of mass energy
$\sqrt{s}$.
The worst behavior comes from the first three diagrams
that grow like $E^2$ while the fourth diagram grows like $E\, m_f$.
To see this one has to use the expression for the
polarization vector for the longitudinal case,
which is given by
\begin{equation}
  \label{eq:20}
  \epsilon_L=(\gamma\beta,\gamma \vec \beta/\beta) \simeq
  \frac{p^\mu}{M} + \mathcal{O} \left(\frac{1}{\gamma^2} \frac{E}{M}\right)\, .
\end{equation}

To determine the coefficients of the high energy behavior [see
eq.~(\ref{eq:3}) below] we cannot use the
approximate expression in the right-hand side of eq.~(\ref{eq:20}) for
all the diagrams
because we would then lose contributions that modify
the $E\, m_f$ terms. Hence, we should employ consistently the
definitions of the left-hand side and expand the result in powers of
$s$, $t$ or $u$. As an example, for the gauge boson $V_a$,
we have
\begin{equation}
  \label{eq:21}
\epsilon_a^L=(\gamma_a\beta_a,\gamma_a \vec \beta_a/\beta_a),\
  \beta_a=\frac{\sqrt{E_a^2-M_a^2}}{E_a},\
  \gamma_a= \frac{1}{\sqrt{1-\beta_a^2}},\
  E_a = \frac{s +M_a^2 -M_b^2}{2 \sqrt{s}},
\end{equation}
and similarly for the other particles. Next we use the kinematics for
the process
\begin{equation}
  \label{eq:42}
f(p_1)+\overline{f}(p_2)\to V_a(p_3) + V_b(p_4)
\end{equation}
to write
\begin{align}
  \label{eq:41}
&p_1=(E_n,0,0,\beta_n E_n),&&
p_2=(E_m,0,0,-\beta_m E_m),\\
&p_3=(E_a,\beta_a E_a \sin\theta,0,\beta_a E_a \cos\theta),&&
p_4=(E_b,-\beta_b E_b \sin\theta,0,-\beta_b E_b \cos\theta),\\
&\epsilon_a^L=(\gamma_a \beta_a,\gamma_a \sin\theta,0, \gamma_a \cos\theta),&&
\epsilon_b^L=(\gamma_b \beta_b,-\gamma_b \sin\theta,0, -\gamma_b \cos\theta).
\end{align}
We then use these expressions to evaluate all the amplitudes. In the
end we substitute $\cos\theta$ in terms of the Mandelstam variable
$t$, through the relation,
\begin{equation}
  \label{eq:43}
  \cos\theta=\frac{t-m_n^2-M_a^2+ 2 E_n E_a}{2 E_n E_a \beta_n \beta_a}\, .
\end{equation}

At this point all the amplitudes are expressed in terms of the
Mandelstam variables and the masses. As the Mandelstam variables are
not independent,
we can still use the relation
\begin{equation}
  \label{eq:44}
  s + t + u = m_n^2+m_m^2+M_a^2+M_b^2
\end{equation}
to express the result in terms of just two independent
variables. Next we want to
isolate the terms that grow with $E^2$ and $E\, m_f$.
To achieve this we introduce the scaling
\begin{equation}
  \label{eq:45}
  s\to s/x, \quad t\to t/x, \quad u\to u/x\, ,
\end{equation}
and make an expansion for small $x$. This would be enough for the
amplitudes without fermions, but here we have the additional
complication of having strings like
\begin{equation}
  \label{eq:2}
  \overline{f}(p_2) \cdots
  f(p_1)\, .
\end{equation}
Since we want to isolate the coefficients of these
structures, and as the spinors grow like $E^{1/2}$,
we also employ the scaling
\begin{equation}
  \label{eq:4}
  \overline{f}(p_2) \cdots  f(p_1)\to \frac{1}{\sqrt{x}}\
\overline{f}(p_2) \cdots f(p_1)\, .
\end{equation}
There is one final complication. Since we wish to have independent spinor
structures, we shall use the Dirac equation between spinors. But then we have
a problem for structures like
\begin{equation}
  \label{eq:11}
  \overline{f}(p_2)\cdots \gamma_\alpha\cdots f(p_1) \epsilon^\alpha\, .
\end{equation}
We have verified that for this case it is enough to use the first term
on the right-hand side of eq.~(\ref{eq:20}).
That is,
we will make the replacement
\begin{equation}
  \label{eq:12}
  \gamma^\alpha \epsilon_{\alpha}(p)\to \frac{1}{\sqrt{x}}\
  \gamma^\alpha\, \frac{p_\alpha}{M}\, .
\end{equation}

The terms that grow as $E^2$ are the
coefficients of $x^{-1}$ and the terms that grow as $E m_f$ are the
coefficients of $x^{-1/2}$.
Therefore we can write for each amplitude
\begin{align}
  \label{eq:3}
  M_i = &\, \overline{f}(p_2)\slashed{p}_3 P_L f(p_1) A_i^L x^{-1} +
  \overline{f}(p_2)\slashed{p}_3 P_R f(p_1) A_i^R x^{-1} \nonumber\\[+2mm]
+ &\, \overline{f}(p_2) P_L f(p_1) B_i^L x^{-1/2}
+ \overline{f}(p_2) P_R f(p_1) B_i^R x^{-1/2}\ +\
\text{constant},
\end{align}
where we have assumed energy-momentum conservation.
We did this consistent
expansion using \texttt{FeynCalc} and \texttt{Mathematica} for the
Lorentz and Dirac algebra and series expansion, respectively.

\subsubsection*{The $E^2$ terms}

The first three diagrams in fig.~\ref{fig:FFAA} yield terms
that grow like $E^2$.
To simplify the expressions, we redefine the coefficients
\be
\hat A_i =A_i M_a M_b\, .
\ee
The corresponding $\hat A_i$ coefficients are given in table~\ref{tab:1}.
\begin{table}[htb]
  \centering
  \begin{tabular}{|c|c|c|}\hline
    Diagram & $\hat A_i^L$ &$\hat A_i^R$  \\[+1mm]
    \hline\hline
    $\mathcal{M}_s^{A}$ & $g_{abe}\ g^L_{emn}$
    &$ g_{abe}\ g^R_{emn} $  \\[+1mm]  \hline
    $\mathcal{M}_t^{f}$ & $ -g^L_{apn}\ g^L_{bmp}$
    &$ -g^R_{apn}\ g^R_{bmp}$  \\[+1mm]  \hline
    $\mathcal{M}_u^{f}$ & $g^L_{amp}\ g^L_{bpn} $
    &$g^R_{amp}\ g^R_{bpn}  $  \\[+1mm]  \hline
    $\mathcal{M}_s^{\phi}$ &  0   & 0  \\[+1mm]  \hline
  \end{tabular}
  \caption{Coefficients $\hat A^L_i$ and $\hat A^R_i$.}
  \label{tab:1}
\end{table}
Since the sum of these coefficients has to vanish,
we end up with two sum rules,
\begin{align}
  \label{eq:14}
 & \sum_p \left[ g^L_{apn}\ g^L_{bmp} - g^L_{amp}\ g^L_{bpn}\right] =
  \sum_e g_{abe}\ g^L_{emn}\, ,\\[+1mm]
 & \sum_p \left[ g^R_{apn}\ g^R_{bmp} - g^R_{amp}\ g^R_{bpn}\right] =
  \sum_e g_{abe}\ g^R_{emn}\, .
\end{align}
These relations are the sum rules in eq. (3.3) of ref~\cite{Gunion:1990kf}.
The cancellation of the terms which grow as $E^2$ is guaranteed by the
gauge group structure of the fermion representations,
as shown by Llewellyn Smith~\cite{LlewellynSmith:1973yud} 
(see also \cite{Cornwall:1974km,Weldon:1984wt}).
So,
these sum rules must hold in any spontaneously broken
gauge theory.

\subsubsection*{The $E$ terms}

Having shown that a spontaneously broken gauge theory assures that the
worst high energy behavior cancels, we move to the terms that grow as
a single power of $E$. Here the gauge invariance of the theory is not sufficient to guarantee
cancellation of the bad high energy behavior, and we obtain constraints on the gauge boson couplings to
scalars. 

For convenience we again define,
\begin{equation}
  \label{eq:22}
  \hat B_i\equiv  M_a  M_b B_i
\end{equation}
The results are summarized in table~\ref{tab:2}.
\begin{table}[b!]
  \centering
  \begin{tabular}{|c|c|c|}\hline
    Diagram & $\hat B_i^L$ &$\hat B_i^R$  \\[+1mm]
    \hline\hline
    $\frac{\mathcal{M}_s^{A}}{g_{abe} \left[\frac{(M_a^2-M_b^2 - M_e^2)
    }{2 M_e^2}\right]}$ & $ g^R_{emn} m_n - g^L_{emn} m_m$
    &$ g^L_{emn} m_n - g^R_{emn} m_m $
    \\[+2mm]  \hline
    $\mathcal{M}_t^{f}$ &$ g^L_{apn} g^R_{bmp} m_p\, -g^L_{apn}  g^L_{bmp} m_m$
    & $ g^R_{apn} g^L_{bmp} m_p\, -g^R_{apn}  g^R_{bmp} m_m$ \\[+2mm]  \hline
    $\mathcal{M}_u^{f}$ &$ g^L_{bpn} g^R_{amp} m_p -g^R_{amp}  g^R_{bpn} m_n$
    & $ g^R_{bpn} g^L_{amp} m_p -g^L_{amp}  g^L_{bpn} m_n$\\[+2mm]  \hline \hline
    $\mathcal{M}_s^{\phi}$ &$ -\frac{1}{2} g_{abk} g^L_{kmn}$
    & $ -\frac{1}{2} g_{abk} g^R_{kmn}$\\[+1mm]  \hline
\hline
  \end{tabular}
  \caption{Coefficients $\hat B_i$. }
  \label{tab:2}
\end{table}
To obtain the sum rule in eq.~(3.4) of Ref.~\cite{Gunion:1990kf},
the sum of the coefficients $\hat B_i^L$ and $\hat B_i^R$ has
to vanish separately.
The follow sum rule is then obtained,

\begin{align}
  \label{eq:7}
&\sum_p\left[ m_p\left( g^R_{bmp}\, g^L_{apn} + g^R_{amp}\, g^L_{bpn}\right)
  - m_m\, g^L_{apn}\, g^L_{bmp} - m_n\, g^R_{amp}\, g^R_{bpn}\right] \nonumber
  \\[+1mm]
&\hskip 20mm
+\sum_e{}'\left[ g_{abe} \left[\frac{M_a^2-M_b^2-M_e^2}{2 M_e^2}\right]
\left( m_n\, g^R_{emn}  - m_m\, g^L_{emn}\right)\right] =\frac{1}{2}\sum_k
 g_{abk} g^L_{kmn}\, ,
\end{align}
where $\sum_e{}'$ means that the sum only runs over massive gauge bosons.
Another sum rule can be obtained from eq.~\eqref{eq:7} with
the substitution $L\leftrightarrow R$.
Eq.~\eqref{eq:7} is similar, but not equal,
to eq.~(3.4) of Ref.~\cite{Gunion:1990kf}.
But we can bring eq.~(\ref{eq:7}) to the form of the sum rule of
Ref.~\cite{Gunion:1990kf} using eq.~(\ref{eq:14}) to write,
\begin{align}
  \label{eq:8}
-\sum_p m_m\, g^L_{apn}\, g^L_{bmp} =& -\sum_p m_m\, g^L_{amp} g^L_{bpn}
-\sum_e g_{abe} m_m\, g^L_{emn}\, ,
\nonumber\\[+1mm]
-\sum_p m_n\, g^R_{amp}\, g^R_{bpn} =& -\sum_p m_n\, g^R_{apn} g^R_{bmp}
+\sum_e g_{abe} m_n\, g^R_{emn}\, .
\end{align}
Now we add the last two equations to obtain
\begin{align}
\label{eq:16}
 \sum_p \left[- m_m\, g^L_{apn}\, g^L_{bmp} - m_n\, g^R_{amp}\, g^R_{bpn}
  \right] =&
\sum_p\left[ - m_m\, g^L_{amp} g^L_{bpn} - m_n\, g^R_{apn} g^R_{bmp} \right]
\nonumber \\
&
+ \sum_e{}' g_{abe} \left( m_n\, g^R_{emn} - m_m\,  g^L_{emn}\right)\, .
\end{align}
Finally,
we substitute eq.~(\ref{eq:16}) into eq.~(\ref{eq:7}),
to obtain
\begin{align}\label{eq:15}
  &\sum_p \left[m_p\left( g^R_{bmp}\, g^L_{apn} + g^R_{amp}\, g^L_{bpn}\right)
  - m_m\, g^L_{amp}\, g^L_{bpn} - m_n\, g^R_{apn}\, g^R_{bmp}\right] \nonumber
  \\[+1mm]
&\hskip 20mm
+\sum_e{}'\left[ g_{abe} \left[\frac{M_a^2-M_b^2+M_e^2}{2 M_e^2}\right]
\left( m_n\, g^R_{emn}  - m_m\, g^L_{emn}\right)\right] =\frac{1}{2}\sum_k
 g_{abk} g^L_{kmn}\, ,
\end{align}
which is precisely the sum rule of eq.~(3.4) of
Ref.~\cite{Gunion:1990kf}.
We also obtain a similar rule by substituting
$L\leftrightarrow R$. The other sum rules in eqs.~(3.5) and (3.6) of
Ref.~\cite{Gunion:1990kf} can
be derived from the above in the same way.

\subsection{$FFV\phi$ Sum Rules}

\subsubsection{The Amplitudes}

\begin{figure}[t!]
\center
\includegraphics[scale=1.1]{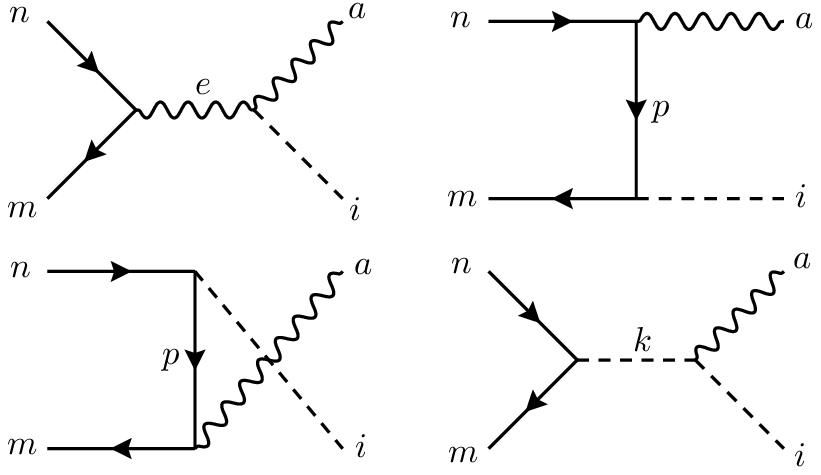}
  \caption{Diagrams for the scattering $f_n + \overline{f}_m \to V_a \phi_i$.}
  \label{fig:FFAS}
\end{figure}

The diagrams contributing to the scattering $f_n(p_1) + \overline{f}_m(p_2) \to
V_a(p_3) + \phi_i(p_4)$
are given in fig.~\ref{fig:FFAS}.
The corresponding amplitudes are given by,
\begin{align}
   \mathcal{M}_s^{A} =&(-i)^2 (i)^2 (g_{aie}) g_{\mu\alpha}
 \overline{f}_m(p_2)\gamma_\nu \left( g^L_{emn} P_L + g^R_{emn}
  P_R\right)\, f_n(p_1)
 \frac{\left[g^{\mu\nu} -
(p_1+p_2)^\mu (p_1+p_2)^\nu/M_e^2 \right]}
{s -M_e^2}
\epsilon^\alpha(p_3)\, ,\nonumber \\
 \mathcal{M}_t^{f} =&(-i) (i)^3\,
\overline{f}_m(p_2) \left( g^L_{imp} P_L + g^R_{imp}
P_R\right) (\slashed{p}_1-\slashed{p}_3 + m_p) \gamma_\alpha \left(
g^L_{apn} P_L + g^R_{apn} P_R\right)  f_n(p_1)
\frac{1}{t -m_p^2} \epsilon^\alpha(p_3)\, ,
\nonumber\\
  \mathcal{M}_u^{f} =&(-i) (i)^3\,
\overline{f}_m(p_2)\gamma_\alpha \left( g^L_{amp} P_L + g^R_{amp}
P_R\right) (\slashed{p}_1-\slashed{p}_3 + m_p) \left(
g^L_{ipn} P_L + g^R_{ipn} P_R\right)  f_n(p_1)
\frac{1}{u -m_p^2} \epsilon^\alpha(p_3)\, ,
\nonumber\\
  \mathcal{M}_s^{\phi} =& (-i) (i)^3 g_{aik}\, \left(-p_4-p_1-p_2\right)_\alpha
\overline{f}_m(p_2)\left( g^L_{kmn} P_L + g^R_{kmn}
  P_R\right)\, f_n(p_1)
     \frac{1}{s-m_k^2}
\epsilon^\alpha(p_3)\, .
\end{align}

\subsubsection{The high energy limit}

In this case there are no divergent $E^2$ terms. The coefficients of
$\hat{B}_i^L$ and $\hat{B}_i^R$ are  summarized in table~\ref{tab:3}.
\begin{table}[b!]
  \centering
  \begin{tabular}{|c|c|c|}\hline
    Diagram & $\hat B_i^L$ &$\hat B_i^R$  \\[+1mm]
    \hline\hline
    $\mathcal{M}_s^{A}$ & $- \frac{1}{2 M_e^2} g_{aei}
                (m_n\, g^R_{emn} - m_m\, g^L_{emn} )$
    &$ - \frac{1}{2 M_e^2} g_{aei}
                (m_n\, g^L_{emn} - m_m\, g^R_{emn} ) $ \\[+2mm]  \hline
    $\mathcal{M}_t^{f}$ & $g^L_{apn}\, g^L_{imp}$
    &$ g^R_{apn}\, g^R_{imp} $\\[+2mm]  \hline
    $\mathcal{M}_u^{f}$ &$-g^L_{ipn}\, g^R_{amp}$
    &$ -g^R_{ipn}\, g^L_{amp} $ \\[+2mm]  \hline \hline
    $\mathcal{M}_s^{\phi}$ &$ g_{aik}\, g^L_{kmn}$
    &$ g_{aik}\, g^R_{kmn} $  \\[+1mm]  \hline
 \hline
  \end{tabular}
  \caption{Coefficients $\hat B^{L,R}_i$. }
  \label{tab:3}
\end{table}
Again we
employed a definition similar to eq.~(\ref{eq:22}),
\begin{equation}
  \label{eq:34}
  \hat B_i^{L,R}\equiv  M_a   B_i^{L,R}
\end{equation}
Since the sum of the coefficients has to vanish,
we obtain the sum rule,
\begin{align}
&  \sum_e{}' \frac{1}{2 M_e^2} g_{aei} (m_n\, g^R_{emn} - m_m\, g^L_{emn} )
- \sum_k  g_{aik}\, g^L_{kmn} =
\sum_p \left( g^L_{apn}\, g^L_{imp}- g^R_{amp}\, g^L_{ipn} \right)\, ,
\end{align}
in agreement with eq. (3.7) of Ref.~\cite{Gunion:1990kf}.
We also obtain a similar sum rule with the interchange $L\leftrightarrow R$.

\end{document}